# Comparison between Networked Control System behaviour based on CAN and Switched Ethernet networks

B. Brahimi, E. Rondeau, C. Aubrun,

*Centre de Recherche en Automatique de Nancy (CRAN-UMR- 7039)*
*Nancy-University, CNRS*
*BP 239*
*54506 Vandoeuvre-lès-Nancy Cedex*
*France*

Abstract: The distributed control systems are more and more used in many industrial applications. These systems are often referred as "Networked control systems". The goal of this paper is to show the network influence on feedback control systems. Two networks are considered: Switched Ethernet network and CAN fieldbus. The first one represents the non-deterministic network and second one represents the deterministic one. Several scenarii are studied to analyse the stability of system according to different network parameters (packets losses, congestion and frame priority). The Truetime simulator is used in this work.

Keywords: Networked control systems, CAN fieldbus, Switched Ethernet network, Simulation.

## 1. INTRODUCTION

The distributed systems are more and more used in many industrial applications in the context of Flexible Manufacturing Systems (FMS), and Reconfigurable Manufacturing Systems (RMS). In the literature, the distributed control systems are referenced under two acronyms ICCS : Integrated Communication and Control Systems [Ray, 1989, Wittenmark, et al. 1995] and NCS : Networked Control Systems, [Zhang, et al. 2001, branicky, et al. 2000, Walsh, et al. 1999, Nilsson 1998, Lian, et al. 2001]. Now, the second acronym is commonly used.

The Networked Control Systems are control systems where controllers, sensors, actuators and other system components communicate over a network. (See figure 1).

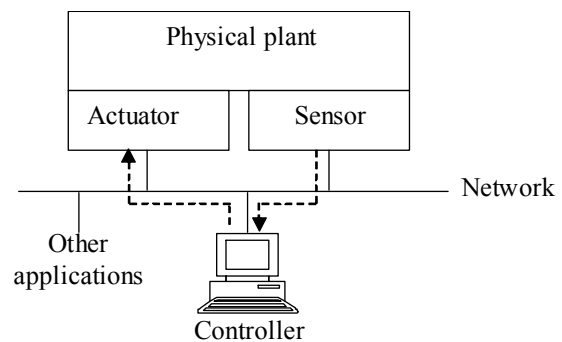

**Figure 1**: Networked Control Systems framework with other applications on the network.

The NCS represents a complex research field due to its pluridisciplinary aspect. NCS studies have to consider simultaneously knowledges on network, computer and control [Lian 2001, Juanole and Mouney 2005].

The insertion of the network communication in the feedback control loop enables to improve, efficiency, flexibility, dependability and modularity of the systems, provides the features to easy installation and reconfiguration and can also reduce the setup and maintenance costs.

However, this has many impacts. The network implementation in the control systems makes the analysis and design of an NCS complex.

Conventional control theories with many ideal assumptions, such as synchronised control and non-delayed sensing and actuation, must be re-evaluated before they can be applied to the NCSs. The network induced delay makes the traditional study of time-delay systems different. Usually, in these systems the delay is considered constant or varying in time interval [lelevé et al 2000]. Another issues caused by the network is the packets losses and the frames scheduling which should be consider during the control system designing. These artefacts can degrade the performance of control systems and can even destabilize the system. These problems are shown in [Halevi and Ray, 1990, Zhang, et al 2001, Juanole and Mouney 2005].

[Wittenmark et al.1995, Sanfridson 2000] discussed several timing issues such as communication and computation delays, processors jitter, and transient errors existing in NCSs. However, only the case where the total maximum network delay is less than one sampling period was considered.

[Bauer, et al. 2001] compensated the network-induced delay by using a controller-based Smith predictor. Nevertheless, in their model they make many ideal assumptions to calculate the network delay such as: the communication between sender and receiver is instantaneous when the communication is available, and the transmission and reception instants are known. These assumptions are not always true (e.g. Switched Ethernet).

[Georges, et al. 2006] compensated the delay by using also the Smith predictor. In this study, the network is a switched Ethernet network. The measurement of the delay is obtained after the device clock synchronisation from the IEEE 1588 protocol.

But the controller-based Smith Predictor model is simple due to a robustness lack. Nowadays this controller is no longer used [Cheong So 2003].

Robust control [Watanabe et al. 1996] has been proposed to compensate the delay effects. [Georges, et al. 2006] compensated the delay by using the robust control and the network induced delay is estimated by using the network calculus theory [Georges, et al. 2005].

In [Walsh et al. 1999], analytic proofs of stability for networked control systems are given for common statistically protocols as well as for a new protocol: Try-Once-Discard (TOD).

[Juanole and Mouney 2005, Juanole, et al 2005] established a relationship between the quality of service of distributed systems and the performances of the closed loop control systems. These works consider the design of distributed systems mechanisms and the control systems in a separate way. Only the fiedbuses are considered.

[Sename et al 2003, Branicky et al. 2003, Branicky et al. 2002, Cervin, et al. 1999, Cervin 2003] defined a new approach in NCS which is the NCS co-design: The design of controller and scheduling is studied in integrated way.

The networks used in the NCSs are often the fielbuses such as: CAN, FIP, ARINC, ControlNet etc… Constructors and academic organisations developed many fieldbuses using deterministic protocols.

Their purpose is to satisfy the real-time requirements. However, the application of fieldbuses has been limited due to the high cost of hardware and the difficulty in interfacing with multivendor products. In order to solve these problems, the computer network technology, especially Ethernet, is being adopted even if the Ethernet protocol is not determinist.

[Lian et al 2001] characterized the network delays for different industrial networks (two fieldbuses: ControlNet and DeviceNet and Ethernet) and studied the inherent tradeoffs between network bandwidth and control sampling rates. They conclude that Ethernet can be used for periodic/non-time-critical and large data size communication, such as communication between or machines. For control systems with short and/or prioritized messages, DeviceNet demonstrates better performance. Nevertheless, the authors were not interested to the Ethernet Switch (Switched Ethernet networks).

[Lee and Lee 02] evaluated both the switched Ethernet performances and the controlled systems. They observed on the networked control system using the switched Ethernet that the control performances were affected very little by the network delay. Therefore, even if an NCS uses a controller designed for the centralized control system without taking into account the network parameters, the NCS with the switched Ethernet network can still satisfy the desired design specifications. Another study [Ji and Kim 05] has shown the feasibility and effectiveness of real-time control system with Ethernet switch.

Recently, the development of switched Ethernet shows a very promising prospect for industrial applications due to the elimination of uncertainties (collisions) in the network operation that leads to improve performance.

The adoption of the switched Ethernet for industrial networking is expected to overcome existing barriers such as high cost, interoperability, and difficulty in application development of the existing fieldbus protocols.

The objective of this paper is to analyse the behaviour of control system regarding two kinds of network: a deterministic network: CAN and a non deterministic network: Ethernet. The interest of this study is to show the importance of the network parameters used in these different networks to control remote applications.

This article is organized as follows: First, The protocols of switched Ethernet and CAN are presented. The second section describes an example of networked control system. The third section studies the network influence on feedback control systems in considered the two previous networks. The results are based on simulations carried out by using TrueTime simulator [Henrikson, et al. 2002, Henrikson, et al.2003]. Finally, conclusions and perspectives are given.

## 2. CAN AND SWITCHED ETHERNET NETWORKS

### 2.1 CAN Network

The Controller Area Network (CAN) serial bus system is used in a broad range of embedded as well as automation control systems. The main CAN application fields include cars, trucks, trains, aircraft, factory automation, industrial machine control, building,…

CAN network [Juanole, et al 2005] [CAN, 2002] is constituted on serial bus shared by stations by mean of a CSMA/CA scheme with a deterministic collision resolution. The collision resolution is based on priorities associated to identifiers (addresses) of the frame which carry the data ($n$ bytes with $n \leq 8$) of application tasks.

The priority (at which message is transmitted compared to another less urgent message) is specified by the identifier of each message. The priorities are laid down during system design in the form of corresponding binary values and cannot be changed dynamically. The identifier with the lowest binary number has the highest priority. Bus access conflicts are resolved by bit-wise arbitration of the identifiers involved by each station observing the bus level bit for bit. This happens in accordance with the wired-and-mechanism, by which the dominant state overwrites the recessive state. All those stations (nodes) with the recessive transmission and dominant observation lose the competition for bus access. All those losers automatically become receivers of the message with the highest priority and do not re-attempt transmission until the bus is available again.

Transmission requests are handled in order of their importance for the system as a whole. This proves especially advantageous in overload situations. Since bus access is prioritized on the basis of the message, it is possible to guarantee low individual latency times in real-time systems.

### 2.2 Switched Ethernet Network

Ethernet was developed in the 1970's and emerged in products in the early 1980s. It is now the dominant local area networking solution in the home and office environment. It is fast, easy to install and the interface ICs are cheap. Despite early attempts to use Ethernet as a real-time communication medium in the factories, practitioners were reluctant to adopt this technology because of its intrinsic non determinism [Decotignie, 2005].

Originally, Ethernet uses a shared medium by using for example hub technology. In this case, simultaneous accesses to the medium generate collisions and the transmission is delayed till no collision occurs. It is means, in the worse case, when the medium is overloaded; a message could never to be transmitted.

Since 1997, new Ethernet versions have been developed and proposed to replace the hub by switches, to connect all the devices in point to point to the switches, to generalise the use of full-duplex mode and to increase the bandwidth. The interest of these technology evolutions is to avoid collisions. But the collision problem is shifted to a congestion problem in switches. The second issue is that the switches generate latencies which have to take into account in control systems.

The use of switches to offer real time guarantees on factory communications has been suggested and analysed by many authors [lee and lee, 2002, Pedreiras et al 2003, Georges, et al, 2005]. The use of switched Ethernet architecture in real-time systems led to develop the IEEE 802.1 D and IEEE 802.1Q. They offer the possibility of compensation and reduction delay with use the prioritisation packet procedure.

### 3. NETWORK INFLUENCES ON FEEDBACK CONTROL SYSTEMS

Many works show the effects of network inside the control loop of the system. Most research in NCS studied the relationships between the network-induced delay and the stability of control systems [Branicky et al. 2003]. And other works [Juanole and Mouney 2005], shown the impacts of Quality of Service (QoS) on the stability of feedback control systems. The QoS mechanisms studied are the frame scheduling, the task scheduling, the drop packets, and the protocols. But, only the fieldbuses are considered.

In this section, the goal is to show the network influence on the closed loop control system. The modelling and simulation are done by using TrueTime simulator [Henrikson et al. 2002, Henrikson, et al.2003]. The considered networks are: switched Ethernet and CAN networks.

Firstly, a servo problem is chosen. The purpose is to follow the command signal. Consider the PD control of a DC servo described by the following continuous-time transfer function:

$$G(s) = \frac{1000}{s(1+s)} \quad (1)$$

Following the specified requirements to have a percentage overshoot less than 5%, $\zeta = 0.7$ the PD parameters are tuned at the following values: $K_p$= 1,5, $K_d$=0.054.
Once the PD controller has been designed in the continuous-time domain and with the appropriate sampling period (h=10ms), we obtained its PD discrete approximation:

$$\begin{aligned} e(t) &= r(t) - y(t) \\ p(t) &= K_p e(t) \\ i(t) &= i(t-h) + \frac{K_i h}{2}(e(t) - e(t-h)) \quad (2) \\ d(t) &= \frac{K_d}{h}(e(t) - e(t-h)) \\ u(t) &= p(t) + i(t) + d(t) \end{aligned}$$

The closed loop control system is distributed on a network (CAN, or switched Ethernet) and modelled with TrueTime simulator.

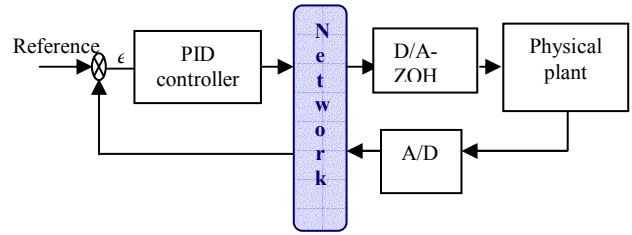

**Figure 2**: Feedback control system model.

### 4. SIMULATION RESULTS

4.1 Introduction
Two kinds of networks: Switched Ethernet and CAN are considered to show the influence of network on feedback control loop. The chosen parameters:

- For the switched Ethernet are: bit rate is 100 Mbit/s, the frame size is 64 Bytes;
- And for the CAN networks: bit rate is 1Mbit/s, the frame size is 8 Bytes.

Three cases are studied:
- Network is used with ideal assumptions,
- Network losses packets (information),
- Network is shared by other applications (Interference disturbances with drop information case).
In all cases, we assume that the congestion don't generate the drop packets. It means that the buffers in network devices are well dimensioning.

4.2 Network used with ideal assumptions

Firstly, the system is simulated in the ideal case. It is means, the network introduces no packets losses and the delay depends only to traffic generated by the real-time system. The network is then not shared with other applications.
The figure 3 shows the behaviour of the system on CAN network and the figure 4 on the switched Ethernet network. In the two cases, the outputs follow the references and the systems are stable. On the CAN experimentations, all the priority configurations of the messages between controller and sensor have been tested. These priority configurations have no impacts on the stability of system.

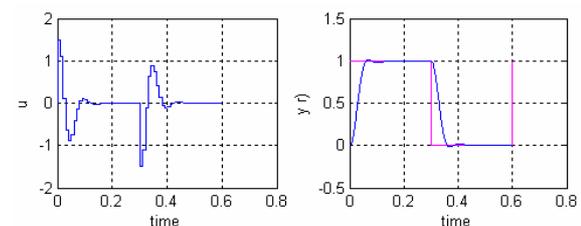

**Figure 3:** Output (y) and controller (u) results of the regulation application on the CAN network.

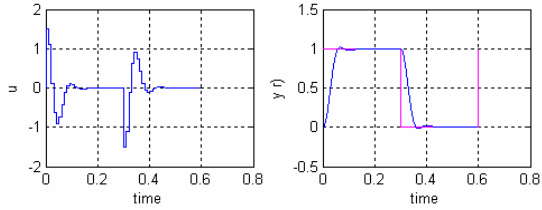

**Figure 4:** Output (y) and controller (u) results of the regulation application on the Switched Ethernet network.

4.3. Information lost by the Network

In this section, three percentages of packet losses are analysed, respectively 5%, 10% and 15%. In the first case, the figure 5 shows a small overshot but the system is stable. And in the second case (figure 6), the system becomes instable, but the system over switched Ethernet is less sensitive to the packet lost than the one over CAN network. And in the last case (figure 7), the system is fully unstable whatever the network used.

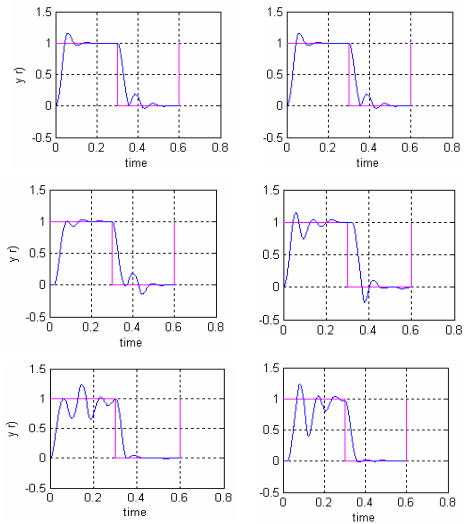

**Figure 5**: Simulation case of networked control system: 1-Switched Ethernet :(In top at left) 5% packets losses, (In middle at left) 10% packets losses (In bottom at left) 15% packets losses.2 -CAN (In top at right) 5% packets losses, (In middle at right) 10% packets losses, (In bottom at right) 15% packets losses.

4.4. Shared network

In this section, the network supports the traffic of other applications. This traffic overloads the medium and this leads to increase the delay of the packets exchanged between the controller and the actuators and sensors of Servo system. The bandwidth occupation of traffic overload is constant, in all evaluated scenarii.
The simulations are analysed with 1%, 10% bandwidth occupation of interference. The period of these perturbations is 7 ms. For CAN network, two cases are evaluated. The first configuration applies a priority 1 (high priority) to the controller and priority 2 to the remote process and the priority 3 to the overload traffic. In the second configuration, the controller has a priority 2, the process the priority 3 and the overload traffic the priority 1. For the switched Ethernet, the scheduling mechanism is FIFO based on store and forward. TrueTime simulator does not implement the classification of service in switch and then it is not possible to associate a priority on an Ethernet frame.

Finally the overload traffic is modelled according two frame sizes: 10 Bytes and 1500 Bytes.

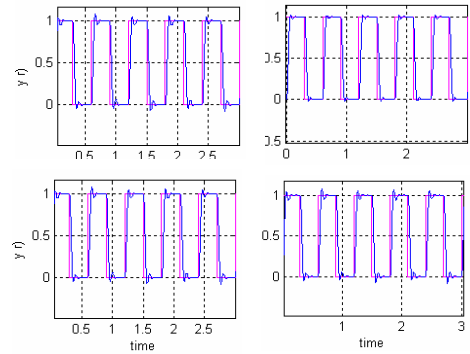

**Figure 6:** Simulation case of networked control system with 10 bytes packet size: 1-Switched Ethernet (At left) 1% bandwidth occupation of interference, 2-CAN (in top at right) 1% of occupation and first priority configuration is applied, (in bottom at right) 1% of occupation and second priority configuration is applied.

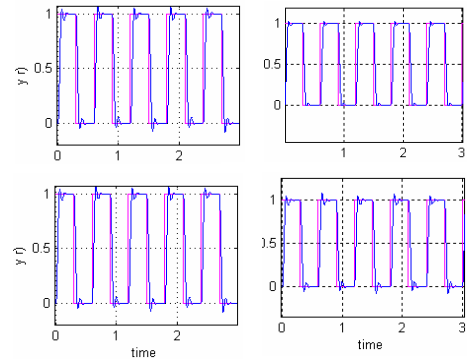

**Figure 7:** Simulation case of networked control system with 10 bytes packet size: 1-Switched Ethernet (At left) 10% bandwidth occupation of interference, 2-CAN: (at the top at right) 10% of occupation and first priority configuration is applied, (in bottom at right) 1% of occupation and second priority configuration is applied

On the CAN experimentations, when the bandwidth occupation of overload is equal to 1% the control system is always stable (both with short and long

packet size). When overload is equal to 10%, the results depend on the priority configurations:
- In case of the overload traffic is tagged with the highest priority, we observe important overshot (> 15%).
- In case of overload traffic is tagged with lowest priority, the system is stable.
- In case of overload traffic is tagged with lowest priority and the packets size is equal to 1500, the system is dramatically instable.

The whole experimentations with switched Ethernet networks shows that the system is always stable

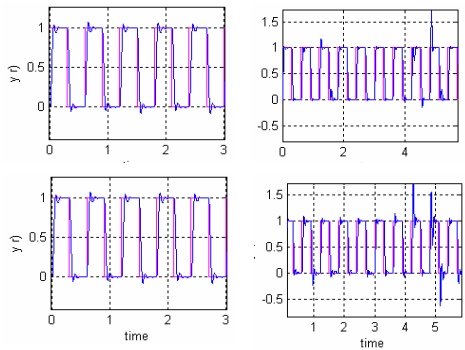

**Figure 8:** Simulation case of networked control system with 1500 bytes packet size: 1-Switched Ethernet (At left) 1% bandwidth occupation of interference, 2-CAN (in top at right) 1% of occupation and first priority configuration is applied, (in bottom at right) 1% of occupation and second priority configuration is applied.

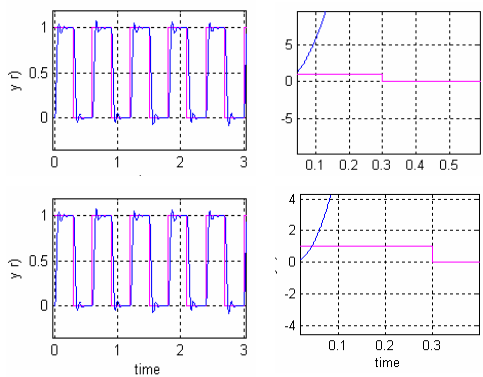

**Figure 9:** Simulation case of networked control system with 1500 bytes packet size: 1-Switched Ethernet (At left) 10% bandwidth occupation of interference, 2-CAN (in top at right) 10% of occupation and first priority configuration is applied, (in bottom at right) 10% of occupation and second priority configuration is applied.

When size of the interference is long, 1500 Bytes (in our case), the switched Ethernet is more efficient (see the figure 8 and 9).

## CONCLUSION

In this paper, we give, at first, an overview on the NCS research field.
And in order to show the impacts of network on the controlled systems, many simulation carried out with TrueTime are realised.
The obtained results show the behaviour of NCS depends on kind of network. Then NCS study has to integrate the specificities of networks such: MAC protocol, frame scheduling, and bandwidth packets size.
The limitation of CAN bandwidth doesn't enable to use it when the real time process shares the same medium with other applications. In this case the use of switched Ethernet architectures seems more suitable (due to high bandwidth)
In the future works we will use the prioritisation packet procedures to compensate the delay due to the insertion of network in closed loop control systems in the framework of switched Ethernet network.

## ACKNOWLEDGMENT

The authors wish to acknowledge the funding support for this research under the European Union 6th Framework Program contract n° IST – 2004-004303 Networked Control Systems Tolerant to faults (NeCST).